\documentclass[prb,showpacs,superscriptaddress,preprintnumbers,amsfonts,amssymb,amsmath,floats,twocolumn,aps]{revtex4}

\usepackage{graphicx}
\usepackage{bm}
\usepackage{color}

\newcommand{\ff}[1]{{\boldsymbol #1}}

\begin{document}

\title{Dynamical mean-field study of partial Kondo screening in the periodic Anderson model on the triangular lattice}

\author{Maximilian W{.} Aulbach}
\affiliation{I. Institut f\"ur Theoretische Physik, Universit\"at Hamburg, Jungiusstra\ss{}e 9, 20355 Hamburg, Germany}

\author{Fakher F{.} Assaad}
\affiliation{Institut f\"ur Theoretische Physik und Astrophysik, Universit\"at W\"urzburg, Am Hubland, 97074 W\"urzburg, Germany}

\author{Michael Potthoff}
\affiliation{I. Institut f\"ur Theoretische Physik, Universit\"at Hamburg, Jungiusstra\ss{}e 9, 20355 Hamburg, Germany}

\begin{abstract}
The competition between Kondo screening and indirect magnetic exchange is studied for a system with geometrical frustration using dynamical mean-field theory (DMFT).
We systematically scan the weak- to strong-coupling regime of the periodic Anderson model on the triangular lattice for a wide range of fillings $n$.
The magnetic phase diagram is derived using a site-dependent DMFT approach by self-consistent mapping onto three independent single-impurity models corresponding to the three correlated $f$ orbitals in the unit cell.
At half-filling, the system is a non-magnetic Kondo insulator for all considered interaction strengths $U>0$ which immediately develops into a non-magnetic metallic Kondo-singlet phase for fillings slightly below half-filling. 
On the other hand, indirect magnetic exchange between the $f$ moments results in antiferromagnetic order at lower fillings.
The antiferromagnetic and the Kondo-singlet phases are separated in the $U$-$n$ phase diagram by an extended region of partial Kondo screening, i.e., a phase where the magnetic moment at one site in the unit cell is Kondo screened while the remaining two are coupled antiferromagnetically.
At even lower fillings the system crosses over from a local-moment to a mixed-valence regime where the minimization of the kinetic energy in a strongly correlated system gives rise to a metallic and partially polarized ferromagnetic state.
\end{abstract}

\pacs{71.27.+a,75.10.Jm,75.30.Mb}


\maketitle

\section{Introduction}\label{sec:intro}

The periodic Anderson model is the model of choice to describe heavy-fermion materials realized in crystals  \cite{Fisk_rev,Tsunetsugu97_rev} or in quantum simulations. \cite{Neumann07,JWerner14a}  
It generically describes a band of light conduction electrons of bandwidth $W$ hybridizing, with matrix element $V$, with a narrow band of $f$ electrons located at energy $\varepsilon_f$.  
Since the $f$ band is narrow, Coulomb correlations are important and are taken into account by an on-site Hubbard interaction $U$. 

The emergent many-body scales depend very much on the choice of bare parameters. 
In the absence of electronic correlations, hybridization leads to an intrinsic $f$-band width $\Gamma = \pi V^2 / W$ which gives an energy scale to assess the impact of the Hubbard $U$. 
In particular, the formation of a local magnetic moment in a metallic host has been studied in Ref.\ \onlinecite{Anderson61}.
It corresponds to a choice of bare parameters where the lower (upper) Hubbard band is below (above) the Fermi energy $\mu$, namely $\varepsilon_f + \Gamma / 2 < \mu <  \varepsilon_f - \Gamma / 2 + U$.
For strong $U$ in the local-moment regime the effective low-energy physics can be approximated by the more simple Kondo-lattice model:\cite{Tsunetsugu97_rev,Capponi00}
Here, charge fluctuations on the $f$ orbitals are completely suppressed, and a super-exchange-like mechanism \cite{SW66} yields a magnetic energy scale, namely a local antiferromagnetic exchange $J  = 8V^{2}/U$ between the local and the  conduction-electron moments.
The local magnetic moment corresponds to a local Kramers doublet and is thereby -- in the absence of correlations -- protected by time-reversal symmetry.    

The residual entropy can be quenched by different and competing mechanisms.
Magnetic ordering breaks time-reversal symmetry and is driven by the Ruderman-Kittel-Kasuya-Yosida (RKKY)\cite{RK54} interaction. 
The corresponding energy scale is set by the effective coupling strength $J_{\rm RKKY}(\ff q) = - J^{2} \chi_s(\ff q,  \omega = 0)$ where $\chi_s$ is the conduction-electron spin susceptibility.
It is an indirect interaction which is mediated via magnetic polarization of the conduction electrons.
This energy scale competes with the Kondo scale \cite{Kondo64,Hewson} given by $T_{\rm K} \propto e^{-W/J}$. 
For temperatures below $T_{\rm K}$ the local magnetic moment is screened through the formation of a many-body entangled spin-singlet state with the conduction-electron spin degrees of freedom.
The competition between RKKY coupling and Kondo screening leads to the famous Doniach diagram \cite{Don77} and to corresponding quantum phase transitions. \cite{Lohneysen_rev}
 
Some heavy-fermion materials, such as CePdAl, are synthesized on frustrated geometries. \cite{Akira08}    
This introduces another energy scale in the problem associated with the release of frustration via a mechanism of partial Kondo screening (PKS). \cite{MNY+10} 
Here, a site-selective Kondo effect alleviates the frustration thus allowing the remnant spins to order magnetically via the  RKKY interaction.  
Such site-dependent screening can occur spontaneously or can reflect chemically different environments in compounds with large unit cells. 

The mechanism of partial Kondo screening has attracted considerable attention in the past. \cite{Ballou91,BFV11}    
The purpose of the present paper is to study the effect in the Anderson lattice beyond the static mean-field approximation. \cite{HUM11,HUM12}
This is achieved by applying a variant of the dynamical mean-field theory (DMFT) \cite{GKK+96} for the Anderson model on the triangular lattice where the different correlated orbitals in the unit cell are treated independently, similar to a real-space DMFT approach. \cite{PN97b}
We will show that the competition between RKKY coupling and Kondo screening, supplemented by lattice frustration leads to a remarkably rich phase diagram including a PKS phase emerging in the local-moment regime at the border between  paramagnetic heavy-fermion and magnetically ordered phases.  

The Anderson lattice has richer physics than the Kondo-lattice model since it allows for charge fluctuations on the $f$ sites.  In conjunction with strong spin-orbit coupling, for example, this naturally leads to the concept of a topological Kondo insulator \cite{Dzero10,JWerner14} which has argued to be realized in SmB$_6$. \cite{Wolgast13,Min14}   
Here, we extend our study beyond the local-moment regime and consider magnetic phases in the mixed-valence regime.
This is realized when the lower (or upper) Hubbard band overlaps with the chemical potential, i.e., if
$\varepsilon_f +  \Gamma / 2 \simeq \mu$.  
In this regime our numerical calculations indicate a completely different physics and predict ferromagnetic order in particular. 
We argue that the latter is reminiscent of itinerant-electron ferromagnetism caused by strong electron correlations in single-band models on frustrated geometries. \cite{VBH+99}

Our study should be understood as a first step only which contributes towards a deeper understanding of the competition between magnetic order and Kondo screening on frustrated lattice geometries:
While the dynamical mean-field theory treats the local, temporal correlations exactly and correctly accounts for the Kondo effect, it also suffers from the simple mean-field-type description of spatial correlations. 
As concerns the effective RKKY interaction, the DMFT does capture its full spatial structure and therewith the corresponding tendencies towards magnetic ordering but the feedback of non-local magnetic correlations on the one-particle Green's function is neglected. 
We expect that this missing feedback will result in a somewhat biased description which probably overestimates the instabilities against magnetic ordering.
Despite this and other possible deficiencies that are characteristic to any mean-field approach, we believe that the non-perturbative and internally consistent physical picture that is provided by the DMFT will serve as an important starting point for future studies, such as cluster and other extensions of the DMFT concept, \cite{cluster} where the effects of short- and long-range correlations are progressively included.
Even this route cannot be expected to provide a final answer, given the complexity of the problem posed by strong correlations in fermionic models on two-dimensional frustrated lattices, and must be supplemented by complementary approaches, such as variational wave functions (see Ref.\ \onlinecite{ipeps} for an example).
However, the phase diagram derived from the site-dependent DMFT approach presented here will in any case provide a useful starting point  and a valuable point of orientation in this general context.

The article is organized as follows. 
The next section introduces the model, the site-dependent DMFT approach and the solver employed here. 
Section \ref{sec:results} presents the numerical results. 
We discuss the DMFT phase diagram and analyze the different phases and mechanisms in detail. 
Conclusions are given in section \ref{sec:con}.

\section{Model and method}\label{sec:methods}

\begin{figure}
\centerline{\includegraphics[width=0.4\textwidth]{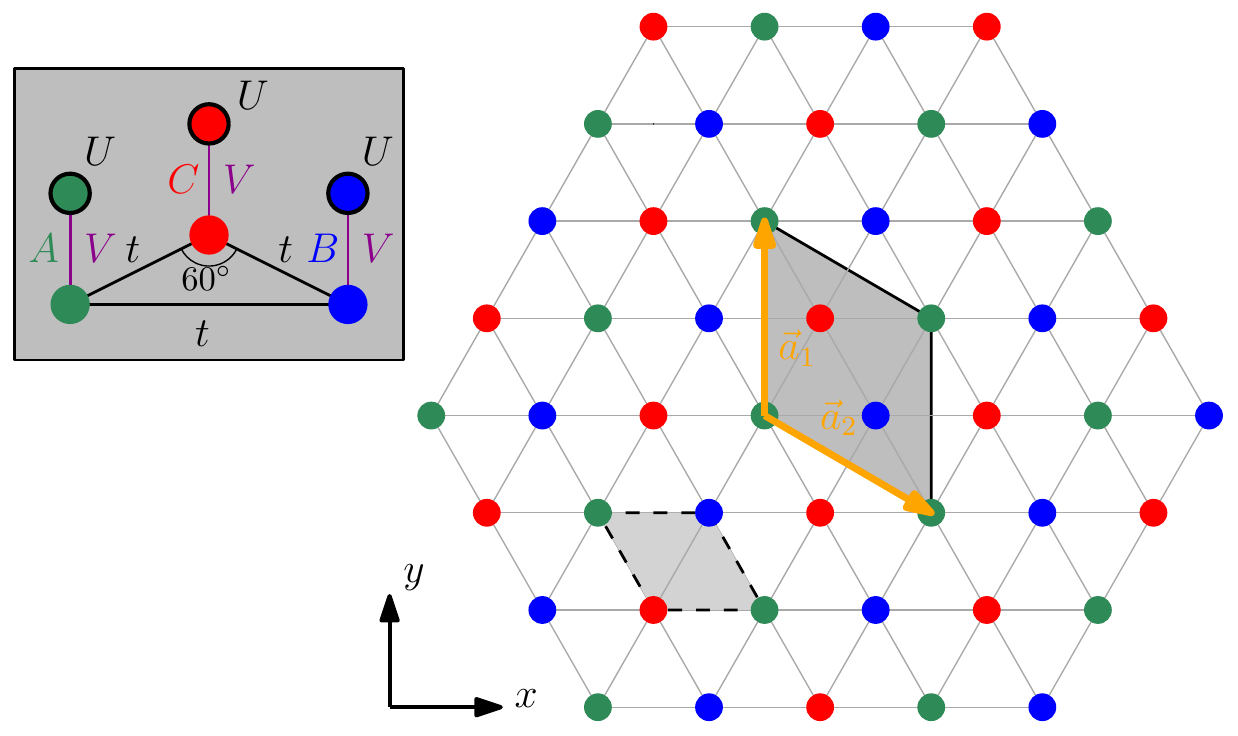} }
\caption{(Color online) 
Periodic Anderson model on the triangular lattice.
{\em Right:}
Primitive unit cell (light gray, dashed lines) and unit cell (gray, solid lines), spanned by the vectors $\mathbf{a}_{1}$ and $\mathbf{a}_{2}$, considered here. 
The latter contains three sites (A, B, C) treated independently within site-dependent dynamical mean-field theory. 
{\em Left:}
For each site, a correlated ($f$) orbital with local interaction $U$ couples to an uncorrelated conduction-electron ($c$) orbital via the hybridization of strength $V$.
The nearest-neighbor hopping $t=1$ between conduction-electron orbitals sets the energy scale.
}
\label{fig:geometry}
\end{figure}

The periodic Anderson model describes correlated ``f'' orbitals with a repulsive on-site interaction which locally hybridize with the ``c'' orbitals of a non-interacting system of itinerant conduction electrons.
We study the Anderson model on the two-dimensional triangular lattice and consider a partitioning of the lattice into non-primitive unit cells containing three sites each, as shown in Fig.\ \ref{fig:geometry}.
Within the variant of the standard DMFT approach employed here, these sites will be treated as inequivalent (see below).
Using standard notations, the Hamiltonian reads:
\begin{eqnarray}
H 
&=& 
\sum_{\mathbf{r} \mathbf{r'}} \sum_{\alpha \alpha'} \sum_{\sigma} c^{\dagger}_{\mathbf{r} \alpha \sigma} t_{\alpha \alpha'}(\mathbf{r} - \mathbf{r'}) c_{\mathbf{r'} \alpha' \sigma}
\nonumber \\
&+&
V \sum_{\mathbf{r} \alpha \sigma} \left( c^{\dagger}_{\mathbf{r} \alpha \sigma} f_{\mathbf{r} \alpha \sigma} + f^{\dagger}_{\mathbf{r} \alpha \sigma}  c_{\mathbf{r} \alpha \sigma} \right)
\nonumber \\
&+&
\varepsilon_{f} \sum_{\mathbf{r} \alpha \sigma} f^{\dagger}_{\mathbf{r} \alpha \sigma} f_{\mathbf{r} \alpha \sigma}
\nonumber \\
&+&
\frac{U}{2} \sum_{\mathbf{r} \alpha \sigma} f^{\dagger}_{\mathbf{r} \alpha \sigma} f_{\mathbf{r} \alpha \sigma} f^{\dagger}_{\mathbf{r} \alpha -\sigma} f_{\mathbf{r} \alpha -\sigma}
\; .
\label{eq:H}
\end{eqnarray}
Here, $\mathbf{r}$ runs over the position vectors to the different unit cells, $\alpha \in \{ {\rm A,B,C} \}$ refers to the sites within a unit cell, and $\sigma\in \{ \uparrow, \downarrow \}$ is the spin projection. 
$c^{\dagger}_{\mathbf{r} \alpha \sigma}$ ($f^{\dagger}_{\mathbf{r} \alpha \sigma}$) creates an electron in the $c$ ($f$) orbital with quantum numbers $\ff r, \alpha, \sigma$. 
Conduction electrons are assumed to hop between nearest-neighboring sites, i.e., the hopping amplitude
$t_{\alpha \alpha'}(\mathbf{r} - \mathbf{r'}) = t \ne 0$ if $\ff r, \alpha$ and $\ff r', \alpha'$ are nearest neighbors.
Furthermore, $V$ is the local hybridization strength, and $U$ is the strength of the Hubbard-type local interaction on the $f$ orbitals.
The one-particle energy of the $f$ orbitals is $\varepsilon_{f}$ and for the $c$ orbitals $\varepsilon_{c} \equiv t_{\alpha \alpha}(0)$.

Dynamical mean-field theory (DMFT) \cite{GKK+96} assumes that the self-energy on the $f$ orbitals be local, $\Sigma_{\ff r \alpha \sigma, \ff r' \alpha' \sigma'}(\omega) = \delta_{\ff r, \ff r'} \delta_{\alpha\alpha'} \Sigma_{\sigma\sigma'}(\omega)$, and maps the lattice problem onto an effective single-impurity Anderson model with one-particle parameters or, equivalently, with a hybridization function $\Delta_{\sigma\sigma'}(\omega)$ that is determined from the local element of the lattice Green's function $G_{\rm loc,\sigma\sigma'}(\omega)$ via the DMFT self-consistency condition. 
This implicitly assumes that the dynamical mean-field $\Delta_{\sigma\sigma'}(\omega)$ is homogeneous. 
Consequently, only homogeneous phases of the DMFT equations can be found in this way. 
In the real-space DMFT approach \cite{PN97b} the self-energy is still assumed as completely local but inhomogeneous solutions of arbitrary complexity are allowed by keeping the full spatial dependence of the local self-energy: $\Sigma_{\ff r \alpha \sigma, \ff r' \alpha'\sigma'}(\omega) = \delta_{\ff r, \ff r'} \delta_{\alpha\alpha'} \Sigma_{\ff r\alpha; \sigma\sigma'}(\omega)$. 
Here, we employ a ``site-dependent DMFT'' by assuming that the local self-energy has possibly different elements on the different sites in a unit cell that is larger than a primitive cell. 
Otherwise, the self-energy is taken as homogeneous: $\Sigma_{\ff r \alpha \sigma, \ff r' \alpha' \sigma'}(\omega) = \delta_{\ff r, \ff r'} \delta_{\alpha\alpha'} \delta_{\sigma\sigma'}\Sigma_{\alpha\sigma} (\omega)$. 
Restricting ourselves to collinear magnetic phases for simplicity, we consider a possibly spin-dependent but spin-diagonal self-energy.

For the above-mentioned partitioning of the triangular lattice, this ansatz for the self-energy means that the periodic Anderson model is self-consistently mapped onto three independent and impurity models with possibly spin-dependent but spin-diagonal one-particle parameters.
The impurity models can be solved independently but are coupled indirectly via the DMFT self-consistency equation.
In particular, we do not impose any further condition on the spatial or spin dependence of $\Sigma_{\alpha\sigma} (\omega)$.
Thereby, we can account for different phases, in particular for collinear magnetic phases, characterized by inhomogeneous order parameters within a unit cell.

Below, we briefly list the main equations of the site-dependent DMFT approach: 
Exploiting the remaining translational symmetry, Fourier transformation of the one-particle term of the Hamiltonian (\ref{eq:H}) provides us with a $6\times 6$ hopping matrix 
\begin{equation}
  \bm{\varepsilon}(\mathbf{k}) =
  \begin{pmatrix}
    \varepsilon_{f} & 0 & 0 & V & 0 & 0 \\
    0 & \varepsilon_{f} & 0 & 0 & V & 0 \\
    0 & 0 & \varepsilon_{f} & 0 & 0 & V \\
    V^{\ast} & 0 & 0 & \varepsilon_{c} & \varepsilon_{\rm AB}(\ff k) & \varepsilon_{\rm AC}(\ff k) \\
    0 & V^{\ast} & 0 & \varepsilon_{\rm AB}^{\ast}(\ff k) & \varepsilon_{c} & \varepsilon_{\rm BC}(\ff k) \\
    0 & 0 & V^{\ast} & \varepsilon_{\rm AC}^{\ast}(\ff k) & \varepsilon_{\rm BC}^{\ast}(\ff k) & \varepsilon_{c}
  \end{pmatrix}
  \label{eq:epsm}
\end{equation}
for each wave vector $\ff k$ in the reduced Brillouin zone.
We have
\begin{eqnarray}
\varepsilon_{\rm AB}(\ff k)
&=& 
t \left[1 + 2 \cos \left( k_{y}/2 \right ) e^{-i \frac{\sqrt{3}}{2} k_{x}} \right]
\; ,
\nonumber \\
\varepsilon_{\rm AC}(\ff k)
&=&
t \left[ 1 + e^{-i \frac{\sqrt{3}}{2} k_{x}} \: e^{-i \frac{1}{2} k_{y}} + e^{-i k_{y}} \right]
\; ,
\nonumber \\
\varepsilon_{\rm BC}(\ff k)
&=& 
t \left[ 1 + e^{-i k_{y}} + e^{i \frac{\sqrt{3}}{2} k_{x}} \: e^{-i \frac{1}{2} k_{y}} \right]
\; .
\label{eq:eps}
\end{eqnarray}
With this, and with a guess for the local but site-dependent $f$ self-energy $\Sigma_{\alpha\sigma}(\omega)$ (for $\alpha \in \{ {\rm A,B,C} \}$) we can start the DMFT self-consistency cycle by calculating the elements of the local lattice Green's function via
\begin{equation}
G_{\mathrm{loc}, \gamma\delta,\sigma}(\omega) 
= 
\frac{1}{L} \sum_{\mathbf{k} \in \mathrm{BZ}} \left[ \frac{ 1 }{ \omega + \mu - \bm{\varepsilon}({\mathbf{k})} - \bm{\Sigma}_{\sigma}(\omega) } \right]_{\gamma\delta}
  \, ,
\label{eq:gloc}
\end{equation}
where $\gamma, \delta$ run over the 6 orbitals in the unit cell and where the $6\times 6$-matrix $\ff \Sigma_{\sigma}(\omega)$ is diagonal and non-zero on the $f$ orbitals only. 
$\mu$ is the chemical potential that is used to fix the total particle density. 
The local Green's function is used to determine the hybridization functions of the three single-impurity Anderson models ($\alpha \in \{ {\rm A,B,C} \}$) as
\begin{equation}
  \Delta_{\alpha\sigma}(\omega) = \omega + \mu - \varepsilon_{f} - \Sigma_{\alpha\sigma}(\omega) - \frac{ 1 }{ G_{\mathrm{loc}, \alpha\alpha,\sigma}(\omega)}
  \, .
  \label{eq:DMFT_Delta}
\end{equation}
Having defined the impurity models, the self-consistency cycle is closed by calculating the self-energy $\Sigma_{\alpha\sigma}(\omega)$ for each impurity model independently. 

The computational bottleneck of the DMFT cycle consists in the solution of the effective impurity problems. 
Here, we use the continuous-time quantum Monte-Carlo method \cite{RSL05,GML+11} based on the hybridization expansion of the action of the respective impurity model \cite{WCM+06} at finite but low temperatures $T$.
Since the interaction term is a density-density Hubbard-type interaction only, it is advantageous to employ the segment-picture variant. \cite{WCM+06, WM06} Following Ref.\ \onlinecite{HPW12}, this allows us to directly measure the impurity self-energy $\Sigma_{\alpha\sigma}(i \omega_n)$ on the fermionic Matsubara frequencies $i\omega_{n}$.

\section{Results}\label{sec:results}

\begin{figure}[b]
\centerline{\includegraphics[width=0.49\textwidth]{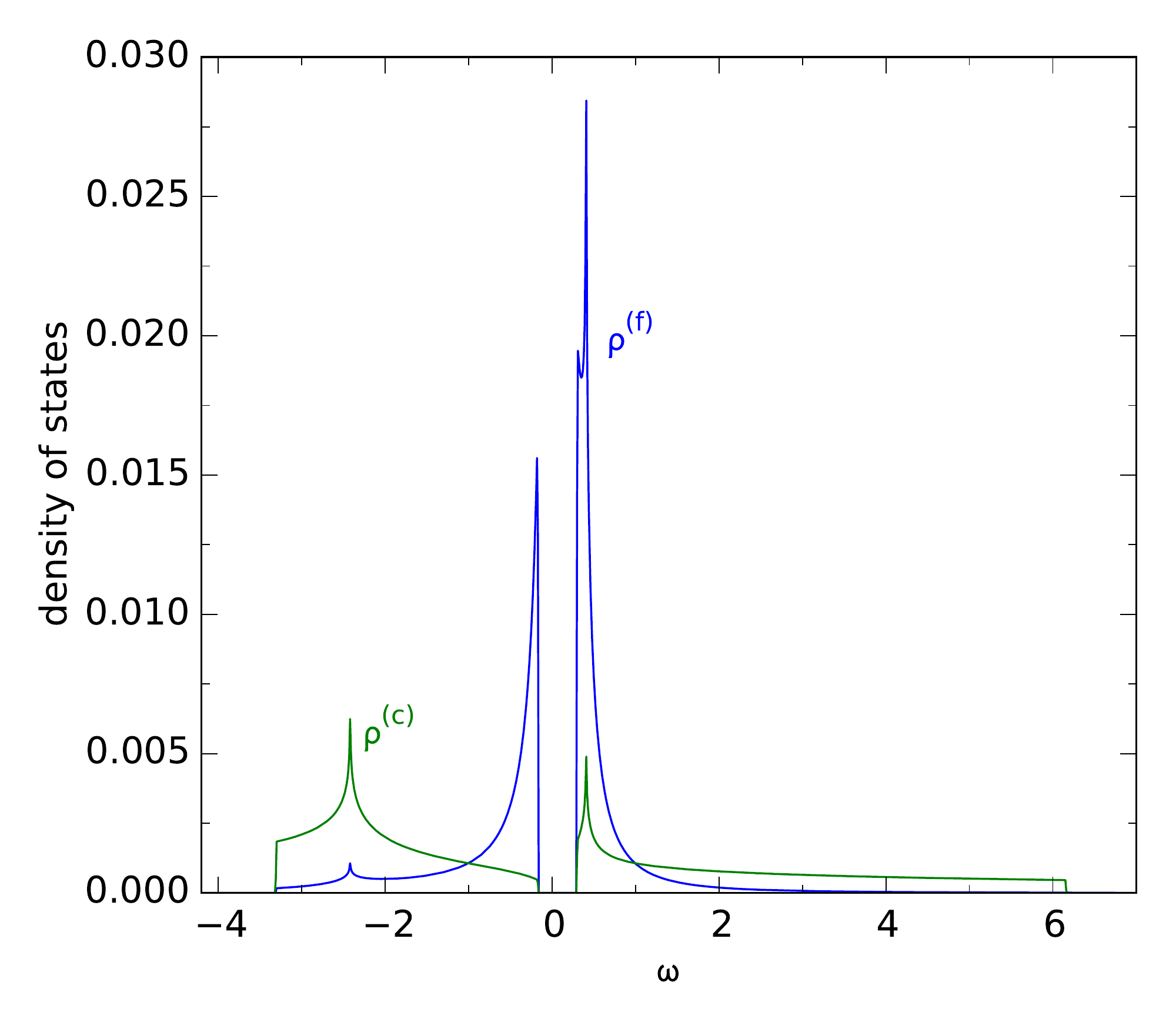}}
\caption{
(Color online) 
$f$- and conduction-electron densities of states $\rho^{(f)}(\omega)$ and $\rho^{(c)}(\omega)$, respectively, for the non-interacting ($U=0$) Anderson model on the triangular lattice. 
Energy units are fixed by the nearest-neighbor hopping $t=1$.
Further parameters: $V=1$, $\varepsilon_{c} = \varepsilon_{f} = 0$.
Centers of gravity are located at $\omega = 0$. 
}
\label{fig:dos}
\end{figure}

DMFT calculations have been performed for the model Eq.\ (\ref{eq:H}) with different chemical potentials $\mu$ to scan the interesting regime at and off half-filling $n=1$ where $n$ is given by
\begin{equation}
n 
= 
\frac{1}{6}
\sum_{\alpha = A,B,C} \sum_{\sigma=\uparrow,\downarrow} 
\left(
n_{\alpha\sigma}^{(f)} + n_{\alpha\sigma}^{(c)} 
\right) 
\: 
\label{eq:filling}
\end{equation}
with 
$n_{\alpha\sigma}^{(f)} = \langle f_{\ff r\alpha\sigma}^{\dagger} f_{\ff r\alpha\sigma} \rangle$ 
and 
$n_{\alpha\sigma}^{(c)} = \langle c_{\ff r\alpha\sigma}^{\dagger} c_{\ff r\alpha\sigma} \rangle$.
The Hubbard interaction is scanned in the weak- to intermediate-coupling range $0 \le U \le 4$ where the nearest-neighbor hopping $t=1$ fixes the energy scale throughout the paper. 
Note that choosing $t>0$ is convenient as this implies that the center of gravity of the total density of states (see Fig.\ \ref{fig:dos}) is located close to the lower band edge. 
Symmetry-broken magnetic phases are therefore expected to occur for fillings below half-filling. 
We furthermore fix the hybridization strength at $V=1$ and choose $\varepsilon_{f} = - U/2$ for the on-site energy of the $f$ orbitals. 
For strong $U$, this ensures that the occupancy of the $f$ orbital at any site $\alpha$ in the unit cell is close to unity, i.e.,
$n^{(f)}_{\alpha} \equiv \langle n^{(f)}_{\alpha\uparrow} \rangle + \langle n^{(f)}_{\alpha\downarrow} \rangle \approx 1$.
The on-site energy of conduction-electron orbitals fixes the energy zero: $\varepsilon_{c} = 0$.

Our main result is the phase diagram for the Anderson model on the triangular lattice as obtained by site-dependent DMFT. 
This displayed in Fig.\ \ref{fig:pd}.
To cover the relevant parameter region, we have performed $\sim 500$ independent DMFT calculations on the SuperMUC supercomputer cluster of the LRZ Munich for different $U$ and $\mu$ in several massively parallel runs with step sizes $\Delta U=0.5$ and $\Delta \mu = 0.05$.
We have considered the model, Eq.\ (\ref{eq:H}) on a lattice with $25\times 25$ unit cells and periodic boundary conditions to perform the $\ff k$-sum in Eq.\ (\ref{eq:gloc}) explicitly.
This is sufficient to ensure that the results do not depend significantly on the system size as has been checked carefully. 
Self-consistent results are indicated as dots and symbols in Fig.\ \ref{fig:pd} in the $U$-$n$ plane. 
About 200 iterations of the DMFT self-consistency cycle usually turned out to be sufficient for convergence. 
To allow for spontaneous breaking of the SU(2) spin-rotation symmetry, we explicitly treat the $\sigma=\uparrow$ and the $\sigma=\downarrow$ channels as independent of each other within the CT-QMC solver. 
Furthermore, the DMFT cycle is started with a spin-asymmetric Hartree-Fock-type initial self-energy. 
It turns out that magnetic phases, if present, are easily found and stabilized in this way.
Within the present study we focus on magnetic phases with collinear moments for simplicity even though non-collinear magnetic phases may be expected in the case of the triangular lattice due to geometrical frustration. 
In fact, previous Hartree-Fock (HF) calculations at and off half-filling \cite{HUM11,HUM12} suggest that a ``classical'' non-collinear $120^{\circ}$ antiferromagnetic phase is realized in a certain range of the phase diagram.
We expect that, by enforcing collinearity, the $120^{\circ}$ phase is replaced by a collinear ``$\uparrow, \uparrow, \downarrow$'' antiferromagnetic phase which has also been found within HF theory. 
\cite{HUM11,HUM12} 

\subsection{Phase diagram}

\begin{figure}[t]
\centerline{\includegraphics[width=0.99\columnwidth]{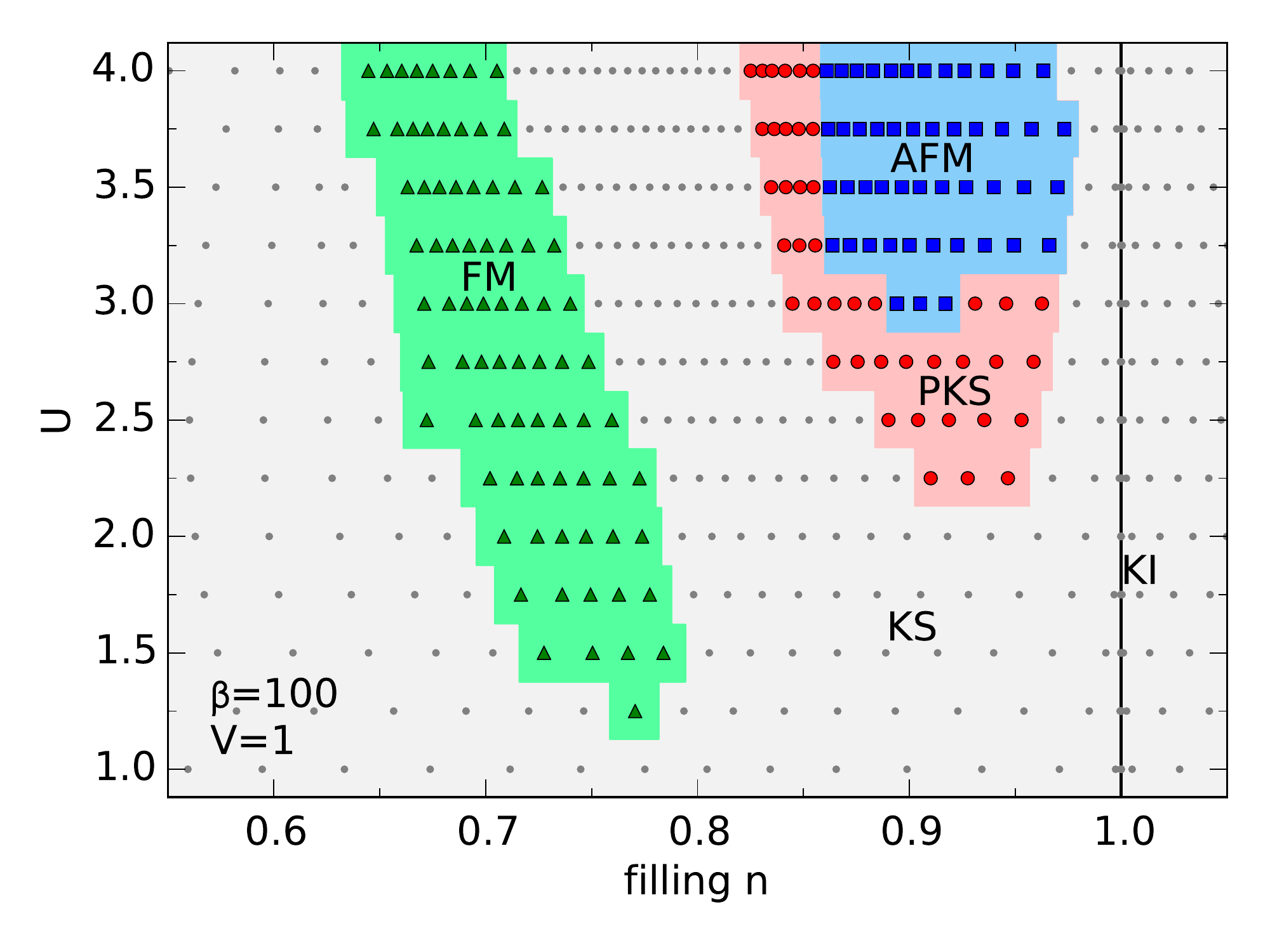} }
\caption{(Color online) 
$U$-vs.-$n$ phase diagram of the Anderson model on the triangular lattice as obtained by site-dependent dynamical mean-field theory. 
Each point corresponds to a converged DMFT calculation using CT-QMC (hybridization expansion, segment code) as a solver at $\beta = 100$.
At half-filling $n=1$ (solid line) the system is a Kondo insulator (KI) for all $U\ge 0$. 
Off half-filling, we find a metallic Kondo singlet state (KS, dots) as well as three different collinear magnetic phases: 
a partial Kondo-singlet phase (PKS, circles), an antiferromagnetic phase (AFM, squares) as well as a ferromagnetic phase (FM, triangles). 
}
\label{fig:pd}
\end{figure}

Fig.\ \ref{fig:pd} shows five different phases. 
At half-filling, the system is a non-magnetic Kondo insulator (KI) in the entire $U$ range considered here. 
For fillings sightly off half-filling, the system stays non-magnetic but immediately becomes metallic. 
Above half-filling, this non-magnetic ``Kondo singlet'' (KS) phase is the only phase that has been found, at least up to $n=1.1 - 1.2$. 
Below half-filling and for a sufficiently strong interaction $U > U_{c} \approx 2$, there are two different magnetic phases, an antiferromagnetic phase (AFM) and a phase with partial Kondo screening (PKS). 
The AFM phase is a collinear ``$\uparrow, \uparrow, \downarrow$'' phase where the magnetic moments at two sites (say, A and B) in the unit cell are ferromagnetically aligned and of equal magnitude while the third moment is antiferromagnetically oriented to the former two with a magnitude such that the total magnetic moment in the unit cell is zero: 
$m^{(f)}_{A} + m^{(c)}_{A} = m^{(f)}_{B} + m^{(c)}_{B} \equiv m_{0} > 0$ and $m^{(f)}_{C} + m^{(c)}_{C} = - 2 m_{0} < 0$. 
Here, $m_{\alpha}^{(f)} \equiv n^{(f)}_{\alpha \uparrow} - n^{(f)}_{\alpha \downarrow}$ and 
$m_{\alpha}^{(c)} \equiv n^{(c)}_{\alpha \uparrow} - n^{(c)}_{\alpha \downarrow}$.

The PKS phase is characterized by one site (say A) with vanishing ordered magnetic moment, or almost vanishing moment (see discussion below), while the moment on the two remaining sites are of equal magnitude but antiferromagnetically aligned: $m^{(f)}_{B} + m^{(c)}_{B} \equiv m_{0} = - ( m^{(f)}_{C} + m^{(c)}_{C}) > 0$. 
The total moment in a unit cell is again zero.
The AFM and the PKS phases appear in a certain filling range $n_{c1}(U) < n_{c2}(U)$ which increases in width with increasing $U$ and which is roughly centered around $n \approx 0.9$. 
The PKS phase appears at weaker $U$ as compared to the AFM phase and separates the latter from the non-magnetic KS phase for $n \to n_{c1}(U)$. 
At much lower fillings, there is also a ferromagnetic phase (FM) with a non-zero total magnetic moment per unit cell. 
This requires a significantly weaker critical interaction $U_{c} \approx 1.25$ as compared to AFM and PKS magnetic phases.
The FM phase is realized in a rather narrow filling range, roughly centered around $n\approx 0.75$ for weak $U$ and $n\approx 0.67$ for $U =4$.

\begin{figure}[b]
\centerline{\includegraphics[width=0.6\columnwidth]{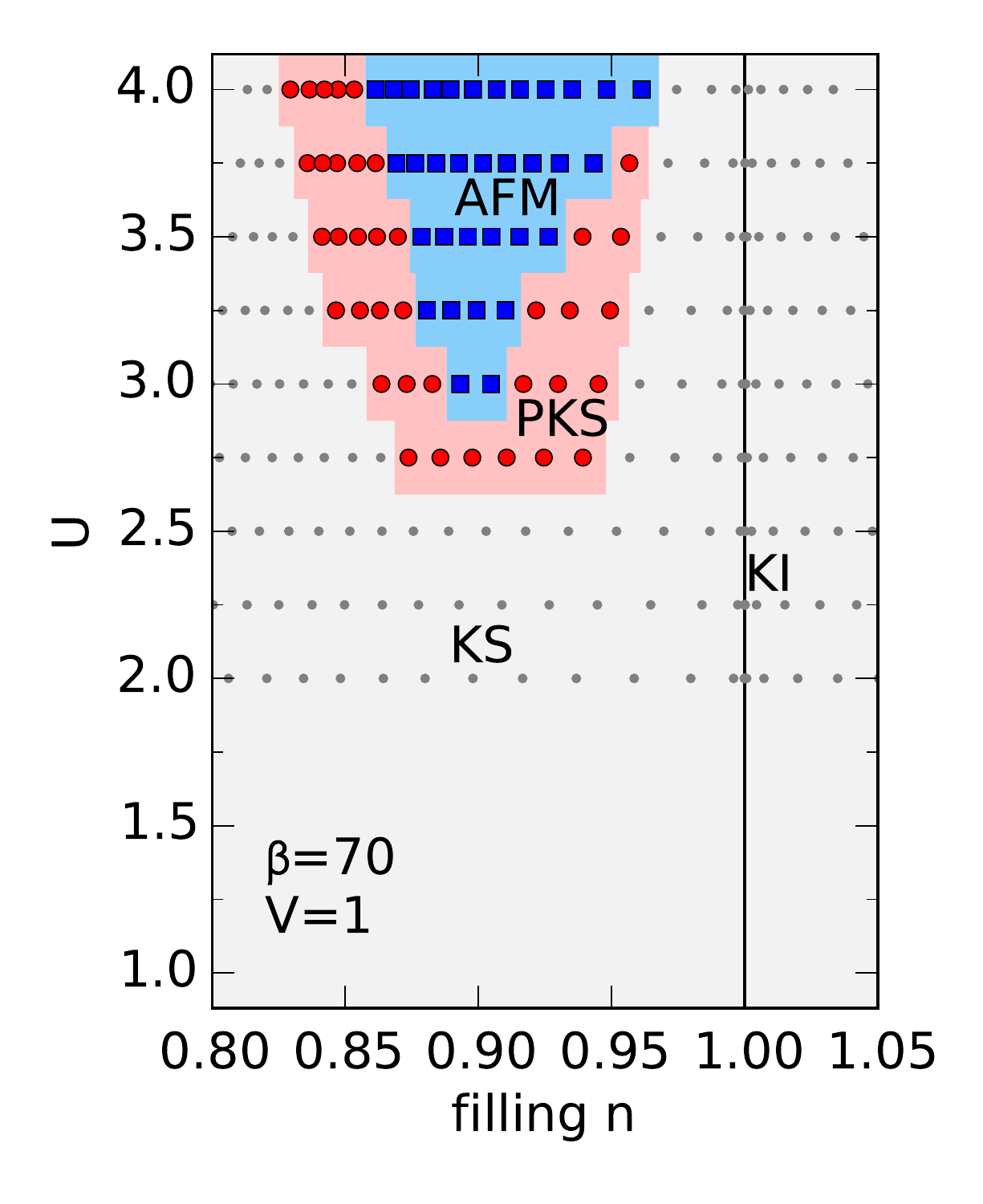} }
\caption{(Color online) 
$U$-vs.-$n$ phase diagram as in Fig.\ \ref{fig:pd} but for $\beta = 70$ and in a smaller filling range including the AFM and PKS phases.
}
\label{fig:pd70}
\end{figure}

We expect that the phase diagram obtained for inverse temperature $\beta=100$ and shown in Fig.\ \ref{fig:pd} is close to the zero-temperature phase diagram. 
To estimate the remaining effects that are due to a finite $\beta$, we have studied the parameter region close to the PKS phase for a somewhat higher temperature ($\beta = 70$). 
The results are shown in Fig.\ \ref{fig:pd70}.
Comparing the phase diagrams for the different temperatures, there are no qualitative differences.
Merely the extension of the AFM and the PKS phases in the $U$-$n$ plane is somewhat reduced for $\beta = 70$, and the critical interaction increases a bit from $U_{c} \approx 2$ ($\beta=100$) to $U_{c} \approx 2.5$ ($\beta=70$).

\subsection{Kondo insulator at half-filling}

\begin{figure}[t]
\centerline{\includegraphics[width=0.45\textwidth]{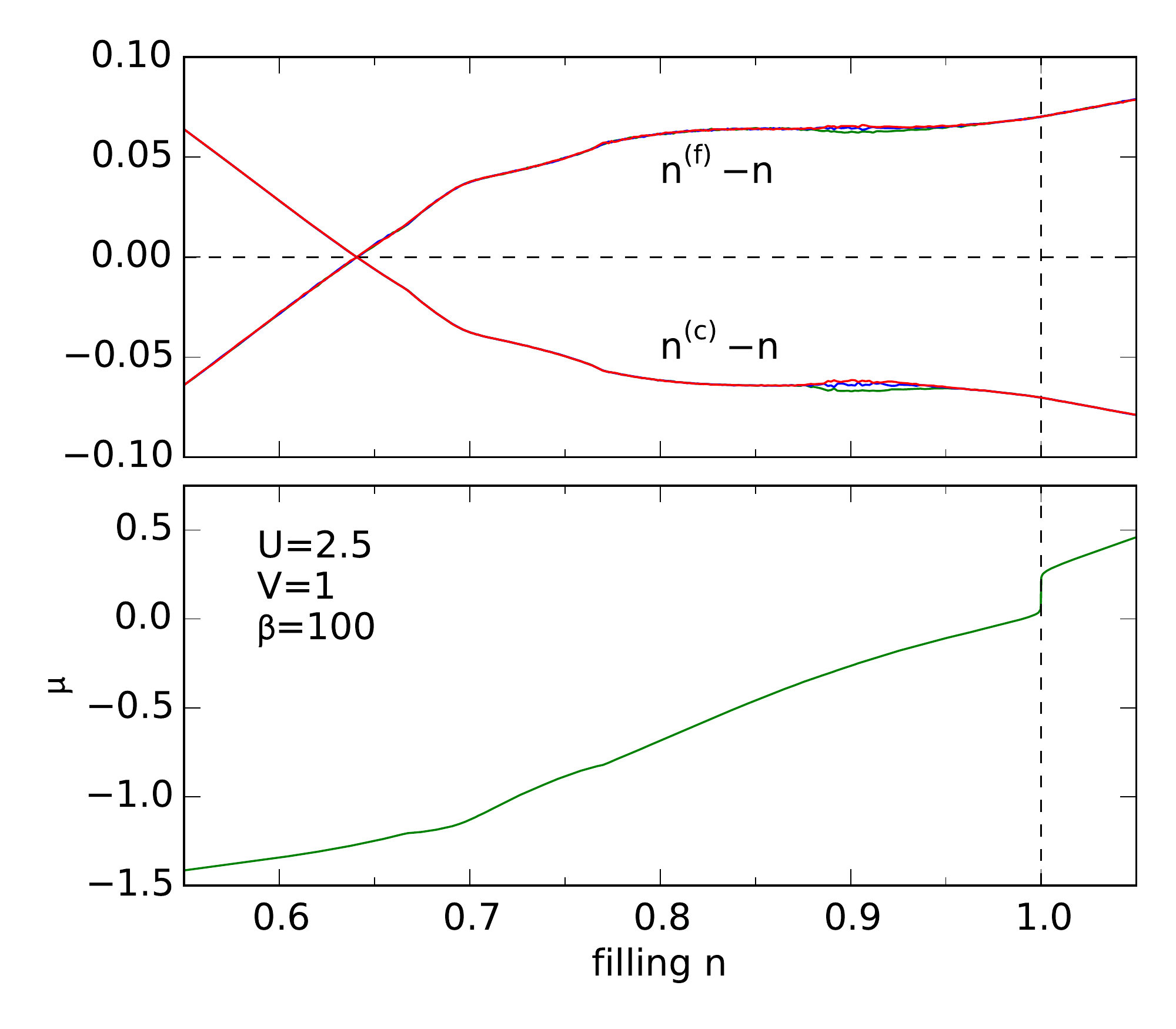}}
\caption{(Color online) 
{\em Upper panel:} 
Difference between the occupancy of the $f$ ($c$) orbitals and the average filling, $n^{(f)}_{\alpha} - n$ ($n^{(c)}_{\alpha} - n$), as function of $n$. 
Results for different sites in the unit cell: A (green), B (blue), C (red). 
{\em Lower panel:}
Chemical potential $\mu$ as function of $n$. 
Calculations for $U = 2.5$. 
Further parameters: $V=1$, $\beta =100$.
}
\label{fig:n25}
\end{figure}

\begin{figure}[b]
\centerline{\includegraphics[width=0.45\textwidth]{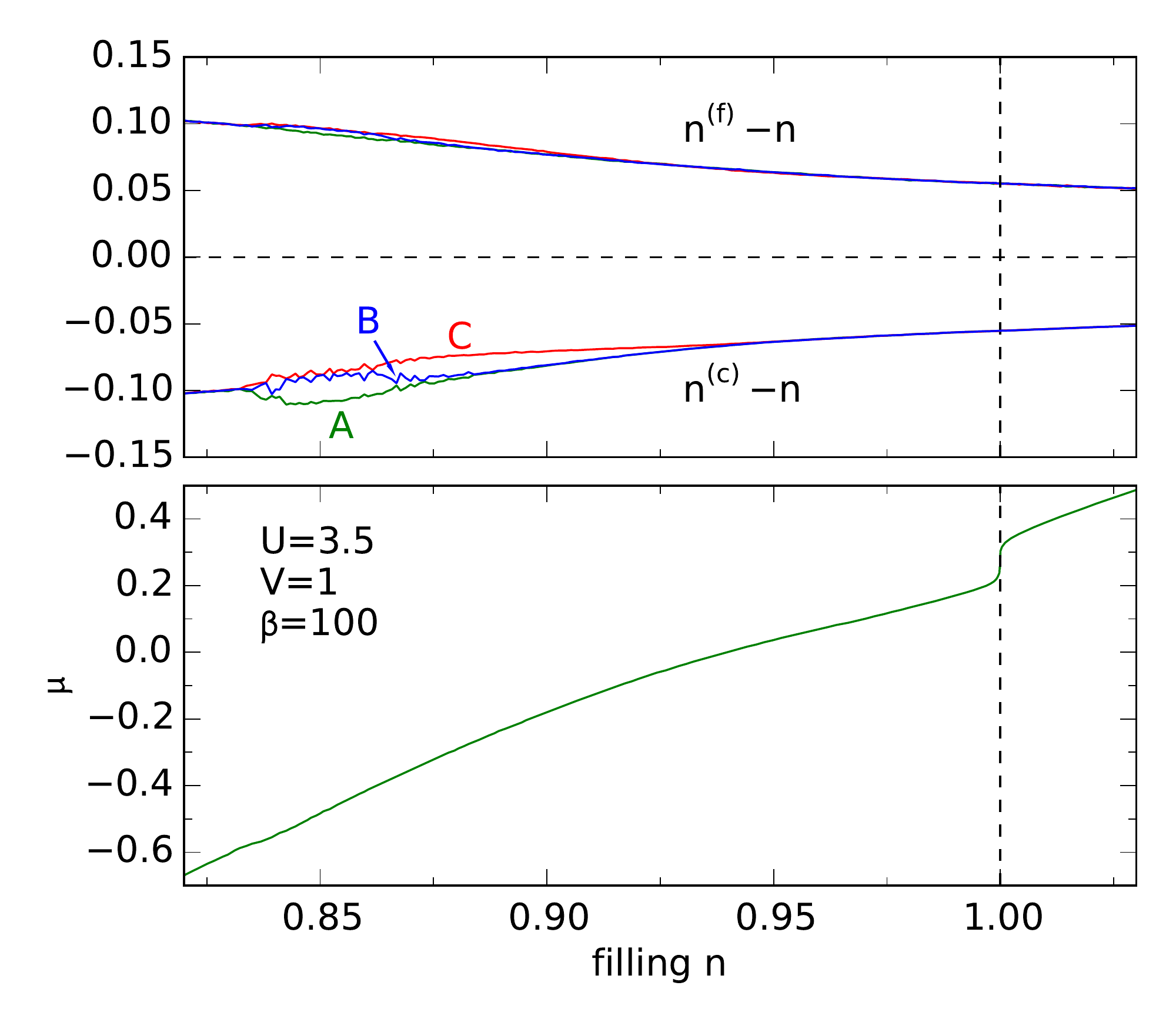}}
\caption{(Color online) 
The same as Fig.\ \ref{fig:n25} but for $U=3.5$.
Results for different sites in the unit cell: A (green), B (blue), C (red), as indicated.
}
\label{fig:n35}
\end{figure}

We start the discussion with the KI phase at half-filling. 
The insulating nature of this phase is easily verified by means of the charge susceptibility $\kappa = \partial n / \partial \mu$ which is found to vanish at half-filling for any $U\ge 0$.
For $U=0$ and half-filling the system is actually a simple band insulator: 
The chemical potential is located in the hybridization  band gap which opens for any $V>0$, see Fig.\ \ref{fig:dos}.
For the correlated system at $U=2.5$, the charge gap $\Delta_{c}$ at half-filling can be read off from the $\mu$ range in which the charge susceptibility $\kappa$ vanishes; see lower panel of Fig.\ \ref{fig:n25} where $\mu$ is plotted as a function of the $n$.
The gap persists for all $U>0$ and decreases with increasing $U$ as is obvious when comparing with the charge gap for $U=3.5$, for example, which can be read off from the lower panel in Fig.\ \ref{fig:n35} (note the different scales for $\mu$ in the two figures). 

It is tempting to relate this decrease of the energy scale with increasing $U$ to the decrease of the coupling constant $J=8V^{2}/U$ in the effective low-energy Kondo lattice that is formally obtained by the Schrieffer-Wolf transformation \cite{SW66} in the local-moment regime of the Anderson lattice model.
Local magnetic moments, required for magnetic long-range order, are formed on the $f$ orbitals due to a strongly repulsive Hubbard-$U$.
One must be aware, however, that even for $U=4$ there are still substantial charge fluctuations on the $f$ orbitals. 
This is indicated, for example, by a $\sim 5\%$ deviation of the average $f$ occupancy from unity at half-filling (see upper panels of Figs.\ \ref{fig:n25} and \ref{fig:n35}). 
Hence, the system is not fully in the local-moment limit. 
Nevertheless, we find an antiferromagnetic linear response of the conduction-electron magnetic moments when applying a homogeneous static magnetic field to the $f$ electron spins.
This indicates an antiferromagnetic ($J>0$) coupling between the local $f$ and $c$ spins consistent with the local-moment picture provided by an effective Kondo lattice.

Deep in the local-moment regime for $U\to \infty$ at fixed $V$, the physics would be governed by a small energy scale, set by $J$, or even by $T_{\rm K} \propto e^{-W/J}$, which makes calculations at stronger $U$ extremely difficult: 
In fact, we have not been able to stabilize a self-consistent solution of the DMFT equations at interaction strengths substantially stronger than $U=4$.

Interestingly, there is no magnetic phase found at half-filling $n=1$. 
This is opposed to static mean-field (HF) theory for the same model \cite{HUM11} which generates a rather complex phase diagram which comprises different magnetic as well as insulating and metallic phases {\em at half-filling}.
Using the Hartree-Fock approximation for the self-energy, 
\begin{equation}
\Sigma^{(f)}_{\alpha\sigma}(\omega) = U \langle n^{(f)}_{\alpha-\sigma} \rangle \: ,
\end{equation}
we have reproduced the HF results of Ref.\ \onlinecite{HUM11} for $V=1$ as a check of our numerical implementation.
As the DMFT correctly accounts for local fluctuations beyond the static mean-field theory, we conclude that those local  fluctuations are sufficient to destroy any magnetic order at $n=1$ (and in the $U$ range considered here).

\subsection{Metallic Kondo singlet phase}

For fillings slightly off half-filling, the system becomes immediately metallic and has a finite charge compressibility $\kappa>0$ (see lower panels of Figs.\ \ref{fig:n25} and \ref{fig:n35}). 
Actually $\kappa$ turns out as non-zero for any filling.
Opposed to previous HF calculations, \cite{HUM12} this implies that there is no instability towards phase separation.

The local correlations between $f$ and $c$ spins are strongly antiferromagnetic as indicated by a corresponding antiferromagnetic linear response.
Still, there is no magnetic order for fillings $n \lesssim 1$.
We refer to this paramagnetic metallic state with local antiferromagnetic correlations as a heavy-fermion or Kondo-singlet state (KS) even if the local spin on the $f$ orbitals, $\ff S_{\alpha}^{(f)} = \frac{1}{2} \sum_{\sigma\sigma'} f_{\alpha\sigma}^{\dagger} \ff \sigma_{\sigma \sigma'} f_{\alpha\sigma'}$, cannot be seen as a rigid spin-$S=1/2$ since the local $f$ moment $(\ff S_{\alpha}^{(f)})^{2}$ somewhat deviates from $S(S+1)=3/4$. 

The respective top panels of Figs.\ \ref{fig:m25} and \ref{fig:m35} show the $f$ orbital double occupancy relative to its non-interacting value, i.e., 
$D_{\alpha} \equiv \langle f_{\alpha \uparrow}^{\dagger} f_{\alpha \uparrow} f^{\dagger}_{\alpha \downarrow} f_{\alpha \downarrow} \rangle / (n^{(f)}_{\alpha \uparrow} n^{(f)}_{\alpha \downarrow})$.
While the double occupancy is suppressed considerably for fillings close to half-filling, it is still far from zero even at $U=3.5$ (Fig.\ \ref{fig:m35}), for example, where $D_{\alpha} \approx 0.45$.
At $U=2.5$ (Fig.\ \ref{fig:m25}) we find $D_{\alpha}$ at a minimum for $n \approx 0.65$.  

\subsection{Antiferromagnetism}

\begin{figure}[t]
\centerline{\includegraphics[width=0.95\columnwidth]{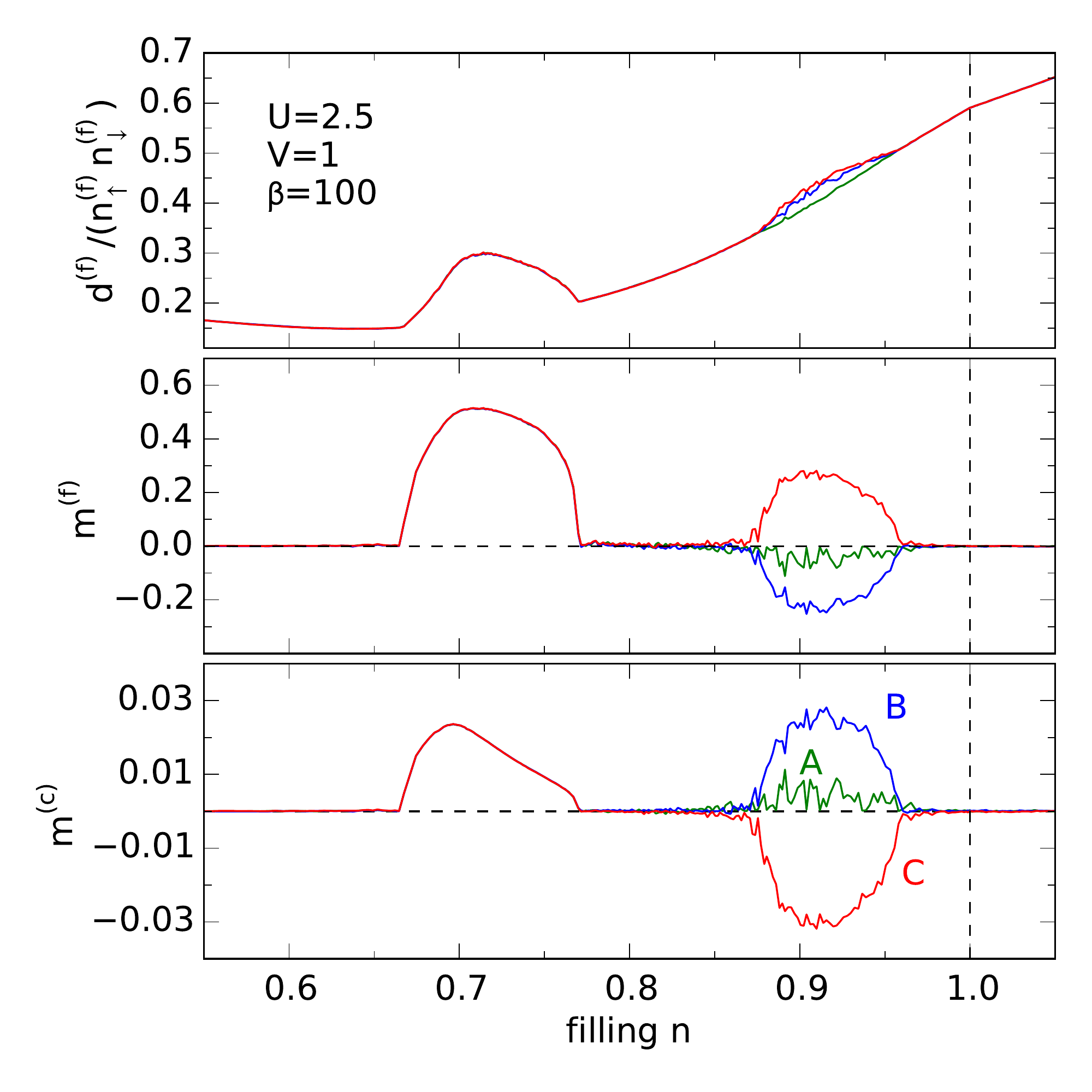} }
\caption{(Color online)
{\em Top panel:}
Filling dependence of the $f$-orbital double occupancy $d_{\alpha}^{(f)} = \langle f_{\alpha \uparrow}^{\dagger} f_{\alpha \uparrow} f^{\dagger}_{\alpha \downarrow} f_{\alpha \downarrow} \rangle$ (relative to its non-interacting value $n^{(f)}_{\alpha \uparrow} n^{(f)}_{\alpha \downarrow}$) for $U=2.5$. Results for different sites A (green), B (blue), C (red) in the unit cell as indicated. 
{\em Middle and bottom panels:}
$f$ and $c$ ordered magnetic moments, $m_{\alpha}^{(f)}$ and $m_{\alpha}^{(c)}$ at the different sites in the unit cell for $U=2.5$ as functions of $n$.
}
\label{fig:m25}
\end{figure}

\begin{figure}[t]
\centerline{\includegraphics[width=0.95\columnwidth]{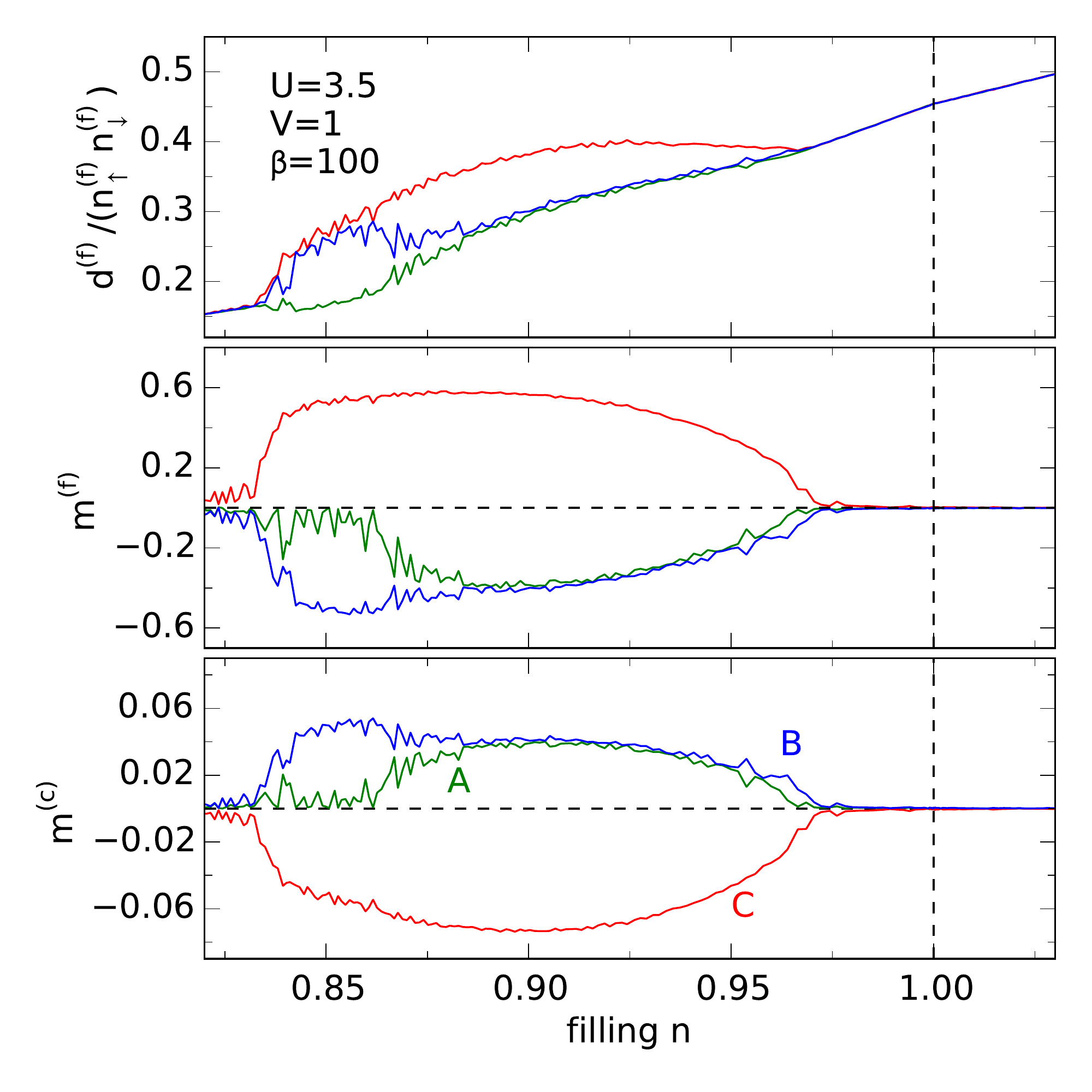} }
\caption{(Color online) 
The same as Fig.\ \ref{fig:m25} but for $U=3.5$ and for a smaller filling range.
}
\label{fig:m35}
\end{figure}

Magnetic phases first appear at fillings centered around $n \approx 0.92$ for $U=2.5$ and for $U=3.5$ (Figs.\ \ref{fig:m25} and \ref{fig:m35})
This is the filling range where the $f$ occupancy is at or very close to unity and where, despite substantial charge fluctuations, the local-moment picture is most adequate.
The magnetic coupling between the local moments must be provided by the {\em a priori} uncorrelated $c$ orbitals, similar to the standard RKKY mechanism \cite{RK54} that can be derived perturbatively in the Kondo-lattice model.

On the triangular lattice, however, magnetic order induced by indirect antiferromagnetic exchange is frustrated. 
Except for a non-magnetic state, there are two obvious possible compromises to form a state with vanishing total magnetic moment in the unit cell, namely a state with $120^{\circ}$ orientations between pairs of magnetic moments as well as a collinear ``$\uparrow, \uparrow, \downarrow$'' phase. 
Apart from the PKS phase to be discussed below, the latter is the only plausible antiferromagnetic state if collinearity between the moments is enforced as is done here.

For $U=3.5$ and with decreasing filling $n$ the system undergoes a phase transition to the AFM phase at $n \approx 0.97$. 
Fig.\ \ref{fig:m35} demonstrates that this phase transition is continuous with the staggered magnetization $m_{0}$ (see definition above) as an order parameter that evolves from $m_{0}=0$ and increases with decreasing $n$ in a continuous way. 
The magnetism is predominantly carried by the $f$ moments with a maximum of $|m^{(f)}_{B}| \approx 0.6$ while the $c$ orbitals are by about one order of magnitude less polarized (note the different scales in Fig.\ \ref{fig:m35}). 
Note that the site-dependent moments are oriented antiparallel to the respective $f$ moments.

Across the transition to the AFM phase there is hardly any change of the double occupancy $\langle f_{\alpha \uparrow}^{\dagger} f_{\alpha \uparrow} f^{\dagger}_{\alpha \downarrow} f_{\alpha \downarrow} \rangle$, i.e., the increase of $D_{\alpha}$ seen in Fig.\ \ref{fig:m35} (top panel) is mainly due to the polarization of the $f$ orbital only. 
For the ``$\downarrow$'' site in the ``$\uparrow, \uparrow, \downarrow$'' state this effect is a bit stronger as its magnetic moment has the higher absolute value.
The fact that the double occupancy and thus the size of the local $f$ moment is basically unaffected, favors a picture of magnetic ordering of preformed local moments and is consistent with an RKKY-like indirect exchange mechanism in the local-moment regime of the Anderson lattice.

\subsection{Partial Kondo screening}

With further decreasing $n$ at $U=3.5$ there is another second-order phase transition from the AFM state to a phase with partial Kondo screening (PKS) (see Fig.\ \ref{fig:m35}). 
For $U=2.5$ the PKS phase directly evolves from the KS through a second-order transition (see Fig.\ \ref{fig:m25}).
In both cases, a Kondo-singlet formed at one site in the unit cell, say A, coexists with a non-local pair of antiferromagnetically coupled moments at the B and C sites. 
The total ordered moment in a unit cell is zero.
Eventually, for fillings $n < n_{c1} \approx 0.88$ at $U=2.5$ and for $n < n_{c1} \approx 0.82$ at $U=3.5$, the system returns to a paramagnetic KS state in another continuous phase transition. 

As compared to the AFM and also, at even lower fillings, to the FM phase, the numerical stabilization of a self-consistent PKS solution is most difficult, i.e., a large number of iterations (up to 200) is required. 
This also reflects itself in the remaining (unphysical) noise on the PKS data seen in Figs.\ \ref{fig:m25} and \ref{fig:m35}.
As a technical remark, let us mention that each DMFT run is completely independent from the preceding one and starts from the same initial guess for the self-energy which is taken as frequency independent and homogeneously spin-polarized. 
Due to this independency, the self-consistent values for the magnetic moments typically do not always form continuous functions of $\mu$, because arbitrary permutations of the A, B, C sites in a unit cell and also a global sign change $\sigma \to -\sigma$ yield physically equivalent solutions of the DMFT equations. 
We have employed those symmetry operations {\em a posteriori} in scans with extremely small steps in the chemical potential ($\Delta \mu = 0.007$) to generate functions as continuous as possible by means of least-square fits minimizing the parametric distance between pairs of consecutive self-consistent solutions. 

The Kondo effect requires a locally antiferromagnetic effective coupling between the local $f$ and $c$ spins. 
This is clearly present: 
As mentioned above, the linear response of the $c$ moments to a static magnetic field applied to the $f$ electron spins is found as antiferromagnetic in paramagnetic phase close to the AFM and PKS phase. 
Furthermore, within the symmetry-broken PKS phase, the ordered moments $m^{(f)}_{\alpha}$ and $m^{(c)}_{\alpha}$ are antiferromagnetically aligned on the B and C sites.
On the other hand, the robustness of the PKS phase, i.e., its extension in the $U-n$ plane, and also the presence of strong charge fluctuations, see the sizable double occupancy in Figs.\ \ref{fig:m25} and \ref{fig:m35}, suggest that the physics is non-universal and by no means ruled by a single Kondo scale $T_{\rm K}$.

It is interesting to note that our data unambiguously show that there is no ``perfect'' partial Kondo screening. 
Namely, a slight polarization $m_{A}^{(f)} < 0$ and $m_{A}^{(c)} > 0$ of the local $f$ and $c$ spins on the A, i.e., on the Kondo site is clearly visible in Figs.\ \ref{fig:m25} and \ref{fig:m35}. 
The proximity to the pair of RKKY-like antiferromagnetically coupled moments, which explicitly breaks time-reversal symmetry, implies that there are admixtures of states with non-zero spin quantum number to the Kondo ``singlet''. 
Assuming this admixture to be given by a single spin-triplet state for simplicity, the antiferromagnetic environment explains a coupling to the $M=0$ component of the triplet. 
A finite polarization of the Kondo singlet, however, rather requires a coupling to the $M = \pm 1$ components and thus implies the additional breaking of the Z$_{2}$ symmetry of the antiferromagnetic state.
This spontaneous symmetry breaking in the PKS phase is also visible in the magnitudes of the B- and C-site moments, namely $|m^{(f)}_{C}| > |m^{(f)}_{B}|$, and is present in the ``$\uparrow, \uparrow, \downarrow$'' AFM state anyway. 

Accompanying the ordering of the spin degrees of freedom, there is a also a (weak) charge ordering in the AFM and the PKS phase (see upper panels of Figs.\ \ref{fig:n25} and \ref{fig:n35}): 
There are two interesting observations: 
First, the deviation of the charge density from the average density is much stronger on the $c$ orbitals as must be expected in the local-moment regime where charge fluctuations on the correlated $f$ orbitals are very effectively suppressed. 
This effect is stronger for $U=3.5$ and compared to $U=2.5$.
Second, within the PKS phase, there is a charge transfer from the ``Kondo site'' (A) to the ``magnetic sites'' (B, C): 
$n^{(c)}_{A} < n^{(c)}_{B,C}$ and $n^{(f)}_{A} < n^{(f)}_{B,C}$.
Due the Kondo effect, the local conduction-electron density of states at the A site will develop a dip, and spectral weight must be shifted above or below the Fermi energy. 
In the absence of particle-hole symmetry, this shift is asymmetric and changes the occupancy. 
The sign and the size of the resulting charge transfer, however, depend on the {\em details} of the band structure. 
Charge disproportionation was also found within the PKS (``partial disorder'') state that is obtained by means of the HF approach. \cite{HUM11,HUM12} 
Opposed to our DMFT results, the charge transfer seen in the HF studies is much larger for the $f$ as compared to the $c$ orbitals. 
This must be seen as an artifact of the static mean-field approach which cannot account for local-moment formation. 

As a function of $U$, the PKS phase is located between the KS and the AFM phase in the phase diagram. 
This can be understood by referring to the famous Doniach diagram: \cite{Don77}
In the KS phase at weaker $U$ (stronger $J$) the Kondo effect dominates while for strong $U$ (weak $J$) the RKKY interaction is dominant and results in magnetic order. 
The PKS state can be seen as a possible way to avoid geometrical frustration in the antiferromagnetically ordered state which is preferred if the formation of a Kondo singlet is less expensive than breaking up two frustrated magnetic bonds and forming a non-frustrated third one. 
As a compromise between indirect exchange, frustration and the Kondo effect, it appears between the KS and the AFM phase.

\subsection{Ferromagnetism}

At lower fillings around $n=0.7$, depending slightly on $U$, the system develops homogeneous ferromagnetic order (see Fig.\ \ref{fig:pd}).
As can bee seen from Fig.\ \ref{fig:m25} for $U=2.5$, the transition to this state is continuous at the lower as well as at the upper critical density.
The ferromagnetic state is metallic with a finite compressibility (see Fig.\ \ref{fig:n25}) and partially polarized with a maximum ordered $f$ moment of $m^{(f)} \approx 0.52$ at $n \approx 0.71$.
The moment on the conduction-electron orbitals ($m^{(c)} \approx 0.02$) is more than an order of magnitude smaller and {\em ferromagnetically} aligned to the moment on the $f$ orbitals.

Generally, there are several mechanisms that may cause metallic ferromagnetism: \cite{Mor85,Cap87,BDN01}
The main idea of the RKKY concept \cite{RK54} consists in a magnetic coupling of well-formed local $f$ moments in an effective Kondo-lattice model \cite{SW66} which is mediated by the conduction electrons and features ferromagnetic order if the effective RKKY coupling $J_{\rm RKKY}(\ff q) = - J^{2} \chi_s(\ff q, \omega = 0)$ is peaked at $\ff q=0$. 
While the RKKY theory is a perturbative approach ($J\to 0$), the double-exchange mechanism \cite{Zen51,deG60,Koc12} applies to the strong-$J$ regime of a Kondo lattice and predicts that a ferromagnetic ordering of the $f$ moments minimizes the kinetic energy of the conduction electrons. 

It is questionable, however, if those concepts apply here as there are strong charge fluctuations preventing the formation of well-defined $f$ moments in our case. 
This is obvious from the sizable deviation of the $f$ occupancy from unity (see Fig.\ \ref{fig:n25}, $n^{(f)} \approx 0.8$ in the relevant filling range).
Another clear indication that the system is no longer in a local-moment regime is the ferro- rather than antiferromagnetic coupling between $f$ and $c$ moments (see middle and lower panel of Fig.\ \ref{fig:m25} around $n=0.7$). 
This is incompatible with an effective low-energy Kondo model. 

It is interesting to note that this implies a filling-dependent crossover from the local-moment regime with a locally antiferromagnetic coupling between $f$ and $c$ moments (see $m_{\alpha}^{(f)}$ and $m_{\alpha}^{(c)}$ in Fig.\ \ref{fig:m25} in the PKS and AFM phases) to a mixed-valence regime. 
This can also be verified easily by studying the linear response in the paramagnetic phase separating the FM and the PKS phase in Fig.\ \ref{fig:pd}:
By applying a weak magnetic field to the $f$ moments, one finds that the local coupling between $f$ and $c$ moments changes from antiferro- to ferromagnetic with decreasing filling.

At $U=0$ the static off-diagonal $f$-$c$ magnetic spin susceptibility can be computed easily in the entire filling range.
Except for low fillings around and below $n\approx 0.25$, corresponding to the van Hove singularity of the density of states close to the lower band edge (see Fig.\ \ref{fig:dos}), the local response is found as antiferromagnetic for $n<1$. 
Above half-filling, the response turns to ferromagnetic and is at a maximum for $n\approx 1.2$ corresponding to the van Hove singularity at $\omega \approx 0.4$ (see Fig.\ \ref{fig:dos}).
We conclude that the ferromagnetic phase cannot be understood as an instability of the Fermi sea in the weak-$U$ regime.
Just the opposite, the paramagnetic state from which the ferromagnetic phase evolves should be considered as strongly correlated. Already for $U=2.5$, the double occupancy is strongly suppressed and $D_{\alpha}$ is in fact at a minimum for $n \approx 0.65$ (see Fig.\ \ref{fig:m25}).

The importance of a strong asymmetry of the density of states for metallic ferromagnetism at strong and intermediate interaction strengths has been emphasized by DMFT studies of the single-band Hubbard model. \cite{Ulm98,PHWN98,VBH+99,PP09}
The key idea is that in a situation where double occupancies are effectively suppressed, the system does not gain much interaction energy from ferromagnetic ordering.
Therefore, the appearance of ferromagnetism must be understood by referring to the (complicated) kinetic energy of the correlated paramagnetic state from which it derives.
Within DMFT this suggests that the {\em shape} of the non-interacting density of states becomes important. 
In fact, studying the impact of a shape-controlling parameter, \cite{WBS+98} ferromagnetism was demonstrated to be favored in cases with a highly {\em asymmetric} density of states, in a parameter range where the density of states is {\em high}, and at strong to intermediate interaction strengths. 

We propose that a similar line of reasoning applies to the periodic Anderson model in the considered parameter region: 
Even at $U=2.5$ and all the more for stronger $U$, double occupancies are strongly suppressed, and the gain in kinetic energy obtained by ferromagnetic ordering is dictated by a strongly asymmetric partial $f$ density of states.
The filling range where ferromagnetism is likely to occur, is then indicated by a corresponding high density of states. 
Note that $n=0.7$ corresponds to a non-interacting chemical potential of $\mu \approx -0.62$ which is already close to the van Hove singularity (at $\omega = - 0.18$).
Substantially higher fillings would be even more favorable for ferromagnetism, but here the crossover to the local-moment regime and the developing antiferromagnetic correlations overwrite the ferromagnetic tendencies.

This picture also explains why the FM phase shifts to lower fillings with increasing $U$ in Fig.\ \ref{fig:pd}: 
Stronger interactions favor ferromagnetism and extend the FM phase to a larger filling range as is again well known from the single-band case. \cite{WBS+98}
This explains the decrease of the lower critical filling for the FM phase with increasing $U$. 
At the same time, however, an increasing $U$ favors local-moment formation, and therefore the KS phase with antiferromagnetic correlations extends at the cost of the mixed-valence regime.
This explains the decrease of the upper critical filling with increasing $U$. 

Previous work \cite{YS93,MW93,TZJF97,DS98,MN00,BBG02} on ferromagnetism in the periodic Anderson model has been done using different theoretical approaches and in largely different parameter regimes. 
Nevertheless, ferromagnetic order away from half-filling appears as a robust result.
As basically all studies have exclusively been performed for bipartite lattices, a direct comparison with our results is not possible. 
There are, however, close similarities with the results of a DMFT study by Meyer and Nolting \cite{MN00} which, for a Bethe lattice with infinite connectivity, demonstrates that ferromagnetism appears in the mixed-valence regime for a finite filling range. 
This study also points out a crossover from antiferro- to ferromagnetic coupling between $f$ and $c$ magnetic moments with decreasing filling, consistent with our findings, and suggests a mechanism based on an effective single-band model with strongly correlated and itinerant electrons -- an idea that was formalized later on by 
Batista et al.\ \cite{BBG02}

\section{Conclusions}\label{sec:con}

Our site-dependent DMFT study of the magnetic phase diagram of the periodic Anderson model on the triangular lattice has uncovered a surprisingly complex phenomenology which could be traced back to a competition between several physical mechanisms at work. 
In particular, the phase diagram is governed by:

(i) the {\em formation of local magnetic moments} on the $f$ orbitals.
Due to the non-bipartite structure of the triangular lattice, half-filling of the $f$ orbitals is found for total fillings below half-filling, around $n \approx 0.9$, weakly depending on $U$.
Here, the low-energy physics is well captured by an effective Kondo lattice although there are sizable $f$ charge fluctuations for the weak- to intermediate-coupling regime considered here ($U\le 4$).
At somewhat lower fillings, there are still well-developed local $f$ moments, but the charge fluctuations increase since the $f$ electrons become itinerant.

(ii) {\em Mixed-valence physics} with strong charge fluctuations on the $f$ orbitals, even at stronger $U$, replaces the local-moment regime for lower fillings (roughly below $n \approx 0.75$, depending on $U$).
The filling-dependent crossover from the local-moment to the mixed-valence regime is accompanied by a reversal of the effective local exchange between the local $f$ and $c$ spins from antiferromagnetic to ferromagnetic. 

(iii) An {\em RKKY-like indirect magnetic exchange} between the $f$ magnetic moments induces antiferromagnetic order for sufficiently strong $U$ within the local-moment regime. 
As we have enforced spin structures to be collinear, this results in an ``$\uparrow,\uparrow,\downarrow$'' AFM phase on the triangular lattice which possibly mimics ``classical'' 120$^{\circ}$ AFM order. 

(iv) The {\em Kondo effect} competes with the indirect exchange in the spirit of the Doniach diagram. 
At low temperatures, besides magnetic ordering, the large entropy carried by the local-moment system can be removed by screening the $f$ moment in a  Kondo singlet with the conduction-electron spin degrees of freedom. 
With decreasing $U$, and prior to charge fluctuations becoming dominant, this Kondo-singlet (KS) phase replaces the antiferromagnetic order.
Kondo physics is also dominating for lower fillings around $n \approx 0.8$, depending slightly on $U$, as well as for fillings very close to and at half-filling. 
The hybridization band gap in the non-interacting density of states results in a band insulator at half-filling for $U=0$ and develops into a correlated Kondo insulator with increasing $U$.

(v) {\em Geometrical frustration} affects the competition between Kondo screening and RKKY coupling. 
At the border between the AFM and KS phase, it becomes favorable to avoid frustration by partial Kondo screening of one $f$ moment per unit cell.
This allows the remnant moments to form an unfrustrated RKKY-coupled collinear antiferromagnet.
The PKS phase is metallic, and it supports a (weak) charge-density-wave ordering in addition, mainly on the $c$ orbitals.
Although it results from a compromise between Kondo screening, RKKY coupling and frustration, the PKS state has turned out to be surprisingly robust.
It appears in an extended parameter range and does not need any anisotropic terms in the Hamiltonian. \cite{MNY+10}
Due to proximity to the time-reversal-symmetry-breaking RKKY-coupled remnant moments, the partial Kondo screening is imperfect resulting in a tiny magnetic moment on the $f$ and, antiferromagnetically aligned, on the $c$ orbital at the ``Kondo site''.

(vi) {\em Strong correlations among itinerant electrons} give rise to a metallic and partially polarized ferromagnetic phase in the mixed-valence regime. 
In this case, the non-bipartite lattice structure favors magnetic order as it produces a highly asymmetric non-interacting density of states which is known to crucially affect the kinetic-energy balance favoring ferromagnetism in a range of fillings with high density of states at the Fermi level and where antiferromagnetic correlations are subdominant.
As a non-perturbative effect, this itinerant-electron ferromagnetism lacks a clear (simple) mechanism -- even in a single-band Hubbard model.

Two main results of our study might be relevant for the understanding of PKS in real materials, such as CePdAl, \cite{Akira08} UNi$_{4}$B \cite{MDN+94} or even artificial geometries of magnetic atoms on metallic surfaces, \cite{Wie09} and for corresponding electronic-structure models:
(i) The PKS state appears at non-integer fillings. 
One might thus speculate that the gain in kinetic energy is essential to stabilize the state and that spin-only models may be questionable. 
(ii) The PKS state exclusively shows up at the border between the paramagnetic heavy-fermion and the magnetically ordered phase. 
This could be tested experimentally by steering the system through this border, either by controlling the temperature or by means of chemical substitution. \cite{FBG+14}

There are several lines along which our study could be continued in the future: 
First, non-collinear phases are in principle accessible by an SU(2)-symmetric formulation of the site-dependent DMFT. 
This may lead to a certain refinement of the magnetic phase diagram, with non-collinear (or even incommensurate) AFM phases partially replacing the ``$\uparrow,\uparrow,\downarrow$'' phase but we do not expect a further qualitative change as the relevant energy scale is still set by the effective RKKY-exchange coupling constant. 

Second, it would be interesting to make contact with the corresponding phase diagram of the Kondo model on the triangular lattice, either by applying DMFT to the Kondo model directly \cite{OKK09} or by using a solver which allows to resolve the Kondo scale, such as the numerical renormalization group. \cite{BCP08}

Finally, one may address the effect of non-local correlations beyond the single-site DMFT. 
The Kondo effect results from correlations between a single correlated $f$ orbital and the conduction-band system and is, therefore, captured correctly by a dynamical mean-field theory which treats those correlations exactly. 
DMFT also provides an accurate description of the non-local indirect exchange but the feedback of non-local magnetic correlations on the self-energy is missing.
Those missing fluctuations must result in mean-field artifacts. 
Typically, the (site-dependent) DMFT approach is therefore, to some extent, biased towards the formation of magnetic order and tends to favor a symmetry-broken state $|\uparrow\rangle |\downarrow\rangle$ at the expense of a non-local singlet 
$(|\uparrow\rangle |\downarrow\rangle - |\downarrow\rangle |\uparrow\rangle)/\sqrt{2}$.
\cite{TSRP12,ATP15}
One might speculate that, compared to the PKS state, the AFM phase is overestimated and that both, the PKS and the AFM phases are overestimated as compared to the KS state. 

\acknowledgments
Financial support of this work by the Deutsche Forschungsgemeinschaft within the Forschergruppe FOR 1346 (projects P1 and P8) and within the Sonderforschungsbereich SFB 668 (project A14) is gratefully acknowledged.
Numerical computations have been performed on the SuperMUC supercomputer cluster of the Leibniz Rechenzentrum M\"unchen.

\end{document}